\documentclass[11pt]{iopart}
\usepackage{iopams}  
\usepackage{graphicx}
\usepackage{amssymb}

\newcommand{\Cset}{{\mathbb C}}
\newcommand{\Pset}{{\mathbb P}}
\newcommand{\be}{\begin{equation}}
\newcommand{\ee}{\end{equation}}
\newcommand{\n}{{\bf n}}
\newcommand{\rr}{{\bf r}}
\newcommand{\ii}{{\mathrm{i}}}
\newcommand{\go}{\mathfrak}                 

\def\Pset{\mathbb{P}}

\def\slashiii#1{\setbox0=\hbox{$#1$}#1\hskip-\wd0\hbox to\wd0{\hss\sl/\/\hss}}
\DeclareGraphicsExtensions{.jpg,.pdf,.gif,.png,.eps}
\begin{document}
\title{Quantum Jumps, EEQT and the Five Platonic Fractals}
\author{A. Jadczyk\dag\  and R. {\"O}berg\ddag }
\address{\dag\ Institute of Theoretical Physics, University of Wroclaw, Pl. Maxa Borna 9, \\50 204 Wroc{\l}aw, Poland}
\address{\ddag\ Senselogic, Olaigatan 2 703 61 \"Orebro Sweden }
\ead{ajad@ift.uni.wroc.pl}
\begin{abstract}
  It is shown that symmetric configurations of fuzzy spin direction
  detectors generate, through quantum jumps, IFS fractals on the sphere
  $S^2.$ The IFS fractals can be also interpreted as resulting from 
  applications of Lorentz boosts to the projective light cone. 
\end{abstract}
\section{Introduction}
Recently, Numerical Algorithms for quantum jumps were discovered.
First they were introduced in quantum optics as convenient and
effective tools for numerical simulations of the Liouville
equation
\cite{carm93,dal92,molmer93,dum92}. A short history and more
references can be found in Ref.
\cite{blaja95a}. Ph. Blanchard and A. Jadczyk, in a series of
papers on EEQT - Event Enhanced Quantum Theory (cf.
\cite{blaja93a,blaja93b,jad94a,jad94b,blaja96b,blaja98a} and
references therein) - developed a new approach that unifies what
John von Neumann called U- and R-processes \cite{vonNeumann} into
a single piecewise deterministic process (PDP) where continuous
evolution of a system is cyclically interrupted by discontinuous
"jumps".

Normally one would expect that time evolution of a physical system
is described by a differential equation. That is how laws of
physics are usually expressed. Here however we have a surprise: in
EEQT a history of an {\em individual quantum system}, coupled to a
monitoring device, as in every real world {\em experiment}, is
described by a {\sl process}\, or an {\sl algorithm}\, rather than
by a differential equation. The PDP is similar to those studied in
the science of economics, where periods of smooth fluctuations are
interrupted by market crashes
\cite{davmha1,davmha2}. A quantum jump is what corresponds to a
market crash - a discontinuity, a 'catastrophe'. But
discontinuities and catastrophes have their own laws, and here
comes the concept of a {\em piecewise deterministic Markov
process\/}\footnote[1]{The Markovian property is not really
important.} - a PDP.

PDPs are the simple and elegant ways to describe the world in
terms of cyclic, rather than linear, time; that is the world of
cyclically, though somewhat irregularly, recurring catastrophes.
In the cases studied in the present paper the catastrophes come
from the coupling of a quantum system to a system of two-state
"detectors". In this case the catastrophes are not really
catastrophes for the detectors - detectors just flip, which is
exactly what detectors do for living. But these flips bring
catastrophes for the quantum system, because with each flip of the
detector, with each "event", as we call it, the quantum system
state vector breaks its continuous evolution, and instantaneously
jumps to a different state - it "rejuvenates" and it starts
another cycle of a peaceful, continuous evolution - till the next
catastrophe.

\noindent{\it Note.}\ The term {\sl instantaneously} in the last
sentence may suggest that our formalism is incompatible with
Einstein's relativity. That this is not so has been demonstrated
in Ref.\cite{blaja96b} 
(for a somewhat different approach cf also \cite{rus02a,rus02b,peres}). 
The point is that for a relativistic
theory the role of time is being played by the Fock-Schwinger
"proper time" as a {\em Floquet\/} variable. The reader should
bear in mind that the EEQT algorithm is explicitly nonlocal: to
simulate a history of an individual system integrations over
entire space (or space-time) are needed.

The EEQT algorithm generating quantum jumps is similar in its
nature to a nonlinear iterated function system (IFS)
\cite{barnsley} (see also \cite{peitgen92} and references therein)
and, as such, it generically produces a chaotic dynamics for the
coupled system. Here the probabilities assigned to the maps are
derived from quantum transition probabilities and thus  depend on
the actual point, but such generalizations of the IFS's have been
also studied (cf.
\cite{peigne} and references therein). In the present paper we
describe the algorithm generating {\em quantum
fractals}, that is self-similar patterns on the projective plane
$\Pset_1(\Cset )\approx S^2$, when a continuous in time
"measurement" of several spin directions at once takes
place.

 {\footnotesize {\bf Note:} The term {\em quantum fractals\/} has been used
before by Casati et al. \cite{casati91,casati99} in a different
context.}

{\footnotesize {\bf Note:} The operators for different spin directions do not
commute, but this does not contradicts Heisenberg's uncertainty
relations as these deal with statistical ensembles averages, while
here we are describing an {\em individual}\/ quantum system. In
fact, realizing the chaotic behavior of a quantum state vector,
when several noncommuting observables are being simultaneously
monitored, can help us to understand the mechanisms of statistical
uncertainties.}

 As stated above the algorithm of EEQT describes a
piecewise deterministic random process - periods of a smooth
evolution interspersed with catastrophic "jumps." Of course, once
we have individual description, we can also get the laws for
statistical ensembles. Here we get a nice, linear, Liouville
evolution equation for measures - as it is usual in studying
chaotic dynamics. The fact that there is a unique PDP generating
the Liouville equation in the framework of EEQT has been proven in
Ref. \cite{jakol95}.

\section{EEQT - Quantum Fractals}
\subsection{Geometry} We have published, as an OpenSource project \cite{perseus},
the algorithm implemented in Java that generates the Five Platonic
Fractals - that is fractals generated by five most symmetric
detector configurations. The algorithm generates self-similar
patterns on a sphere of a unit radius. The points on the sphere
represent (pure) states of the simplest quantum system - the spin
$1/2$ rotator.  This spin $1/2$ quantum system is coupled,
continuously in time, to a finite number of symmetrically
distributed spin-direction detectors. Thus the symmetry of the
pattern reflects the symmetry of the detector directions
distribution. Each spin direction is characterized by a vector
${\bf n}$ of unit length. Here we study the most symmetrical
configurations, therefore we chose direction vectors ${\bf n}_i$
pointing from the origin to the vertices of one of the five
platonic solids. We consider the following five detectors
configurations:
\begin{enumerate}
  \item tetrahedron: 4 detectors along the directions $\n[i], i=1,\ldots ,4$\\
  \{\{0, 0, 1.\}, \{a[17], 0, -a[3]\}, \{-a[6],
    a[12], -a[3]\},
    \{-a[6], -a[12], -a[3]\}\}
  \item octahedron: 6 detectors along the directions $\n[i], i=1,\ldots ,6$\\
  \{\{0, 0, 1.\}, \{1., 0, 0\}, \{0, 1., 0\},\\
   \{-1., 0, 0\}, \{0, -1., 0\}, \{0, 0, -1.\}\}
  \item cube: 8 detectors along the directions $\n[i], i=1,\ldots ,8$\\
  \{\{0, 0, 1.\}, \{a[17], 0, a[3]\}, \{-a[6], a[12],
    a[3]\},
     \{-a[6], -a[12], a[3]\},\\ \{a[6],
    a[12], -a[3]\}, \{a[6], -a[12], -a[3]\},
     \{-a[17],
    0, -a[3]\}, \{0, 0, -1.\}\}
  \item icosahedron: 12 detectors along the directions $\n[i], i=1,\ldots ,12$\\
  \{\{0, 0, 1.\}, \{0.a[15], 0, a[5]\}, \{a[2], a[13],
    a[5]\},
     \{-a[10], a[7], a[5]\},\\ \{-a[10], -a[7],
    a[5]\},  \{a[2], -a[13], a[5]\},
     \{a[10],
    a[7], -a[5]\},\\ \{a[10], -a[7], -a[5]\},
     \{-a[2],
    a[13], -a[5]\}, \{-a[15],
    0, -a[5]\}, \\\{-a[2], -a[13], -a[5]\},
     \{0, 0, -1.\}\}
  \item dodecahedron: 20 detectors along the directions $\n[i], i=1,\ldots ,20$\\
  \{\{0, 0, 1.\}, \{a[9], 0, a[11]\}, \{-a[3], a[8],
    a[11]\}, \{-a[3], -a[8], a[11]\},\\ \{a[11], a[8],
    a[3]\}, \{a[11], -a[8], a[3]\}, \{-a[14], a[4],
    a[3]\},\\ \{a[1], a[16], a[3]\}, \{a[1], -a[16],
    a[3]\}, \{-a[14], -a[4], a[3]\},\\ \{a[14],
    a[4], -a[3]\}, \{a[14], -a[4], -a[3]\}, \{-a[11],
    a[8], -a[3]\},\\ \{-a[1],
    a[16], -a[3]\}, \{-a[1], -a[16], -a[3]\}, \{-a[11], \
-a[8], -a[3]\},\\ \{a[3],
    a[8], -a[11]\}, \{a[3], -a[8], -a[11]\}, \{-a[9],
    0, -a[11]\},\\ \{0, 0, -1.\}\}
\end{enumerate}
where the array of real numbers $a[i],\/ i=1,\ldots ,17$ is given
in the following table.
\begin{center}
\begin{tabular}{|l|l|l|l|}
\hline
$a[1]=\frac{3-\sqrt{5}}{6}$&$a[2]=\frac{5-\sqrt{5}}{10}$&$a[3]=\frac{1}{3}$&$a[4]=\frac{\sqrt{5}-1}{2\sqrt{3}}$\\
\hline
$a[5]=\frac{1}{\sqrt{5}}$&$a[6]=\frac{\sqrt{2}}{3}$&$a[7]=\sqrt{\frac{5-\sqrt{5}}{10}}$&$a[8]=\frac{1}{\sqrt{3}}$\\
\hline
$a[9]=\frac{2}{3}$& $a[10]=\frac{5+\sqrt{5}}{10}
$&$a[11]=\frac{\sqrt{5}}{3}$&$a[12]=\sqrt{\frac{2}{3}}$\\
\hline
$a[13]=\sqrt{\frac{5+\sqrt{5}}{10}}$&
$a[14]=\frac{3+\sqrt{5}}{6}$& $a[15]=\frac{2}{\sqrt{5}}$&
$a[16]=\sqrt{\frac{3+\sqrt{5}}{6}}$\\
\hline
$a[17]=\frac{2\sqrt{2}}{3}$&&&\\
\hline
\end{tabular}
\end{center}
{\footnotesize {\bf Note:} There is no deep reason why we chose
these configurations - simplicity and beauty are the main factors
here. Notice that, to enable easy zooming onto the attractor, we
have chosen the orientations in such a way that in each case the
North Pole, with coordinates (0,0,1), of the sphere is occupied by
one of the vertices.}
\subsection{The algorithm}
Here we describe the algorithm. Comments on its meaning and on
the derivation can be found in the endnotes in Section \ref{sec:notes}
\subsubsection{The Hilbert space}
The quantum system is represented in a two-dimensional complex
Hilbert space, which we realize as $\Cset^2$ - the set of all
column vectors
\be
a=\pmatrix{  a_1 \cr
  a_2}
\ee
where $a_1$ and $a_2$ are complex numbers, and with scalar
product $(a,b)$ defined by
\be
(a,b)={\bar a}_1 b_1+{\bar a}_2 b_2
\ee
where the bar  ${\bar c}$ stands for the complex conjugation of
the complex number $c.$
\subsubsection{Spin directions} We choose the Pauli matrices
$\sigma_x,\sigma_y,\sigma_z$ to represent spin directions along
$x,y,z$ axes respectively.
$$
\sigma_1=\sigma_x=\pmatrix{0,&1\cr 1,&0},\quad 
\sigma_2=\sigma_y=\pmatrix{0,&-\ii\cr \ii,&0},\quad
\sigma_3=\sigma_z=\pmatrix{1,&0\cr 0,&-1}
$$
Together with the identity matrix
\be
\sigma_0=I=\pmatrix{1,&0\cr 0,&1}
\ee
they span the whole $2\times 2$ complex matrix algebra. In
computations it is important to make use of the fact that Pauli
matrices (after multiplication by "-i") represent the quaternion
algebra, that is:
\begin{eqnarray}
\sigma_1^2=\sigma_2^2=\sigma_3^2=I\nonumber\\
 \sigma_1\sigma_2=-\sigma_2\sigma_1=\ii\sigma_3,\ \sigma_2\sigma_3=-\sigma_3\sigma_2=\ii \sigma_1,\
 \sigma_3\sigma_1=-\sigma_1\sigma_3=\ii \sigma_2\nonumber
\end{eqnarray}
To each direction $\n$ in space there is associated spin matrix
\be
\sigma(\n )=n_1\sigma_1+n_2\sigma_2+n_3\sigma_3=\pmatrix{n_3,&n_1-\ii n_2\cr
n_1+\ii n_2,&-n_3}
\ee
satisfying automatically $\sigma(\n )^2=I,$ and with eigenvalues
$+1,-1$. Vectors $\pmatrix{1\cr 0}$ and $\pmatrix{0\cr 1}$ are
eigenvectors of $\sigma_3$ to eigenvalues $+1$ and $-1$
respectively and thus correspond to "North" and "South" spin
orientations respectively. Let $P(\n )$ denote the projection
operator that projects onto eigenstate of $\sigma(\n )$ to the
eigenvalue $+1 .$ Then $P(\n )$ is given by the formula:
\be
P(\n )=\frac{1}{2}(I+\sigma(\n)).
\label{eq:pan}
\ee
Indeed, $P(\n )$ is Hermitian and has eigenvalues
$\frac{1}{2}(1\pm 1)=$ $1$ or $0$ - thus it is the orthogonal
projection, and it projects onto the eigenstate of $\sigma(\n)$
with spin direction $\n .$
\subsubsection{Fuzzy projections}
For each $\n$ let $P(\n ,\epsilon )$ be the fuzzy projection
opearator defined by the formula:
\be
P(\n ,\epsilon )=\frac{1}{2}(I+\epsilon\sigma (\n ))
\label{eq:pna}
\ee
where $0\leq\epsilon\leq 1$ ), or better: $1-\epsilon ,$ is a
parameter that measures the "fuzziness." The extreme cases are
not the very interesting ones: for $\epsilon=0$ we get the identity
operator - maximal fuzziness and no information whatsoever, while
for $\epsilon=1$ we get the sharp projection $P(\n )= P(\n ,\epsilon=1
).$

We restrict the range of the parameter $\epsilon$ to the interval
$[0,1]$ because only in this range $P(\n,\epsilon )$ is a positive
operator. It is easy to see that this is so. Indeed, a Hermitian
matrix is positive when its eigenvalues are positive, and the
eigenvalues of $P(\n,\epsilon )$ are $(1\pm
\epsilon)/2 ,$ thus $-1\leq \epsilon\leq 1 .$
On the other hand negative $\epsilon$ for $\n$ is the same as
positive $\epsilon$ for $-\n ,$ thus we restrict the range of
$\epsilon$ to $[0,1].$

It is the operators $P(\n ,\epsilon )$ that will act on quantum
states to implement "quantum jumps" whenever detectors "flip."

The overall coefficient in the definition (\ref{eq:pan}), chosen
to be $\frac{1}{2}$ here, is not important because in applications
each of the operators $P(\n ,
\epsilon )$ is multiplied by a coupling constant, and, in our case,
when we are not interested in timing of the jumps, the value of
the coupling constant plays no role.
\subsubsection{Jumps are implemented by fuzzy projections}
Let us now discuss the algebraic operation that is associated with
each quantum jump. Suppose before the jump the state of the
quantum system is described by a projection operator $P(\rr)$,
$\rr$ being a unit vector on the sphere. That is, suppose, before
the detector flip, the spin "has" direction $\rr$. Now, suppose
the detector $P(\n ,\epsilon )$ flips, and the spin right after the
flip has some other direction, $\rr^\prime$. What is the relation
between $\rr$ and $\rr^\prime$? It is easy to see that the action
of the operator $P(\n ,\epsilon )$ on a quantum state vector is
given, in terms of operators, by the formula:
\be
\lambda(\epsilon,\n,\rr) P(\rr^\prime) =P(\n ,\epsilon )P(\rr )P(\n ,\epsilon ),
\label{eq:lambda1}
\ee
where $\lambda(\epsilon,\n,\rr)$ is a positive number. It is a
simple (though somewhat lengthy) matrix computation that leads to
the following result:

\be
\lambda(\epsilon,\n,\rr)=\frac{1+\epsilon^2+2\epsilon (\n\cdot\rr )}{4}
\label{eq:lambda2}
\ee
\be
\rr^\prime=\frac{(1-\epsilon^2)\rr+2\epsilon(1+\epsilon(\n\cdot\rr ))\n}{1+\epsilon^2+2\epsilon
(\n\cdot\rr
)}\label{eq:jump}
\ee
where $(\n\cdot\rr )$ denotes the scalar product
$
\n\cdot\rr=n_1r_1+n_2 r_2+n_3 r_3.$

\subsubsection{Transition probabilities}
Given the actual state $\rr$ of the quantum system, and the
configuration of the detectors $\{\n[i], i=1,2,\ldots ,N\}$ we
compute probabilities $p[i], \sum_{i=1}^N p[i]=1$ for the $i-th$
detector to flip. Then we select randomly, with the calculated
probability distribution, the flipping detector, and we implement
the jump by changing $\rr$ to $\rr^\prime$ according to the
formula (\ref{eq:lambda1}), with $\n=\n [i]$. The probabilities
$p[i]$ are computed using the theory of piecewise deterministic
Markov processes applied to the case of quantum measurements - as
developed within EEQT.\footnote{ It is of interest that Born's
probabilistic interpretation of quantum mechanics as well as the
standard formula for quantum mechanical transition probabilities,
can be derived in this way and there is no need of adding it as a
separate postulate. }According to EEQT the probabilities $p[i]$
are given by the formula:
\be
p[i]=const\cdot\mbox{Tr}\left(P(\rr)P(\n[i],\epsilon)^2P(\rr)\right)
\label{eq:prob}
\ee
where $const$ is the normalizing constant. Using cyclic
permutation under the trace, as well as the fact that $P(\rr
)^2=P(\rr )$ we find, taking trace of both sides of the formula
(\ref{eq:lambda1}), that $p[i]$ are proportional to
$\lambda(\epsilon,\n[i],\rr)$ given by (\ref{eq:lambda2}), thus
\be
p[i]=\frac{1+\epsilon^2+2\epsilon (\n[i]\cdot \rr)}{N(1+\epsilon^2)}.
\ee
Note that, owing to the fact that $\sum_{k=1}^N \n[k]={\bf 0}$ we
have $\sum_{i=1}^N p[i]=1$, as it should be.
\section{The Five Platonic Fractals}
As we noted above, different values of $\epsilon$ give different
fuzziness. To produce representative pictures, one for each solid,
we adjusted $\epsilon$ so that, as a rule, the more vertices, the
higher value of $\epsilon$ - thus higher resolution of details. While
rendering the pictures, for obtaining grayscale value for a given
pixel we were using either the formula $\log (data+1)$ or, to
get more details at the peak values, even $log(log(data+1)+1).$ 
\subsection{Other Polyhedra}
In principle our algorithm should create quantum fractals for each
of the regular polyhedra. The only restriction on the array of
vectors $\n[i]$ is that they are all of unit length, and their sum
is a zero vector. We added, for comparison with the Platonic
solids configurations, two additional simple yet regular figures:
double tetrahedron and icosidodecahedron. Notice that tetrahedron
is self-dual, while dodecahedron and icosahedron are dual to each
other. Double tetrahedron array is obtained by combining $\n[i]$
with $-\n[i]$ - that is with the inverted configuration.

Icosidodecahedron has particularly simple and elegant expression
for its 30 vertices: they are of the form: $(\pm 1,0,0)$ and its
cyclic permutations, and $\frac{1}{2}(\pm 1,\phi ,\frac{\pm
1}{\phi})$ and its cyclic permutations, where
$\phi=\frac{1+\sqrt{5}}{2}=1.61803\ldots $ is the golden ratio.
All of its edges are of length $\phi$. For its 30 vertices
$\epsilon=0.85$ was needed to resolve the atrractor's fine
structure.
\section{Notes \label{sec:notes}}
\subsubsection{Complex projective plane serves as a canvas\label{sec:canvas}}
Pure states of the quantum spin (we are discussing spin $1/2$
here) are described by unit vectors in our Hilbert space
$\Cset^2$. But, as it is standard in quantum theory, proportional
vectors describe the same state - the overall phase of the vector
has no physical significance. Therefore, in geometrical terms, the
set of all pure states is nothing but the projective complex space
$\Pset_1(\Cset)$ which happens to be the same as the sphere $S^2.$
That is why the fractal pattern, in our case, is being drawn on a
spherical canvas. There is, however, another possible
interpretation of the same algorithm. It is well know that there
is an intrinsic relation between Minkowski space and the space of
$2\times 2$ complex Hermitean matrices. \footnote{In fact, by
using Cayley transform, this relation identifies the space of
unitary matrices with the compactified Minkowski space.} In
coordinates the map is given by
$
p=\{p^\mu\}\mapsto {\slashiii p}\doteq p^\mu\sigma_\mu ,$
so that $\mathrm{det}({\slashiii p})=p^2=p^\mu p_\mu .$ In
particular our projection operators $P(\rr )$, $\rr^2=1$
correspond to null directions. In other words our canvas, the
sphere $S^2$, can be also thought of as the projective light cone
- the space of light directions. Quantum jumps would then
correspond to sudden changes of directions of light or light-like
entities. Indeed the operators $P(\n,\epsilon )$ are positive, thus
proportional to Lorentz boosts. The formula (\ref{eq:jump}) for jumps 
implemented by these operators should be compared with the formula
for a Lorentz boost, with velocity $\beta$ in the direction $\n$:
$$y^0=x^0 \cosh \alpha +({\bf x}\n)\sinh \alpha,\quad
{\bf y}={\bf x}-[({\bf x}\n)\cosh\alpha + x^0\sinh\alpha ]\n,$$
where $\beta = \tanh \alpha$ is the velocity. It follows that the jumps 
described by the equation (\ref{eq:jump}) can indeed be interpreted as 
Lorentz boosts witht velocity $\beta=2\epsilon/(1+\epsilon^2).$

The formula (\ref{eq:lambda1}) can be easily generalized
for $P(\rr )$ being a generic Hermitean matrix (thus representing
a four-vector $p$ of space- or time-like character as well.
However there is no such generalization for the formula
(\ref{eq:prob}), so that the probabilities would have to be
assigned equal, and a physical interpretation, even tentative one,
is missing in such a case (cf Subsection \ref{sec:hyper} below).
\subsection{Pure states as projection operators}
For a spin $1/2$ quantum system pure quantum states are uniquely
described by projection operators $P(\n )$, where $\n$ is a unit
length direction vector, starting at the origin, and ending on one
of the points of the unit sphere. All pure states are of this
form. Indeed, every orthogonal projection, except of the two
trivial ones: $0$ and $I,$ are of the form $P(\n )$ for some $\n
.$
\subsection{Fuzzy projections}
Detecting the particle spin is somewhat different than detecting
its position components. To measure the position of a particle we
can use a photographic plate or a bubble chamber. In such a case a
simple mathematical model of a detector is obtained by associating
with each active center of the detector a fuzzy, bell-type
function with its half-width corresponding to the active region of
the center ( for instance about $1\mu m$ for an AgBr grain of a
photographic emulsion). What would correspond to such a "fuzzy"
projection operator in the case of a spin measurement? We do not
have much of a choice here. Due to symmetry reasons there is only
one formula possible, namely one given by Eq. (\ref{eq:pna}).
\subsection{Importance of fuzziness}
It is importance to note that in our generalization of the
projection postulate, as the result of a jump, not all of the old
state is forgotten. The new states depends, to some degree, on the
old state. Here EEQT differs in an essential way from the naive
von Neumann projection postulate of quantum theory. The parameter
$\epsilon$ becomes important. If $\epsilon=1$ - the case where $P(\n
,\epsilon)=P(\n )$ is a projection operator - the new state, after
the jump, is always the same, it does not matter what was the
state before the jump. There is no memory of the previous state,
no "learning" is possible, no "lesson" is taken. This kind of a
"projection postulate" was rightly criticized in physical
literature as being contradictory to the real world events,
contradicting, for instance, the experiments when we take
photographs of elementary particles tracks. But when $\epsilon$ is
just close to the value $1$, but smaller than $1,$ the
contradiction disappears. This has been demonstrated in the cloud
chamber model \cite{jad94b}, where particles leave tracks much like
in real life, and that happens because the multiplication operator
by a Gaussian function does not kill the information about the
momentum content of the original wave function. Notice that our fuzzy projection operators 
$P(\n,\epsilon)$ have the properties similar to those of Gaussian
functions, namely\footnote{This can also be interpreted as Lorentz formula for addition of relativistic velocities- cf. Section \ref{sec:canvas} below.}
\be
P(\n,\epsilon)^2=\frac{1+\epsilon^2}{2}P(\n,\frac{2\epsilon}{1+\epsilon^2}).
\label{eq:pn2}
\ee
\subsection{Geometrical meaning of the parameter $\epsilon$}
It is instructive to have a visual picture of the map
$\rr\mapsto\rr^\prime$ of the sphere $S^2$ implemented by the
operator $P(\n ,\epsilon ).$ To this end let us assume the vector
$\n$ is pointing North, i.e. $\n=(0,0,1)$. Then the result
$\rr^\prime$ of applying the operator $P(\n ,\epsilon )$ to a point
$\rr$ on the sphere is on the same longitude as the original point
$\rr$, but its latitude $\theta$ changes - it moves towards the
North Pole along its meridian, the new latitude being given by
the formula:

\be
\theta^\prime = \arccos\left (\frac{(1 - \epsilon^2)\cos(\theta) +
          2\epsilon(1 + \epsilon\cos(\theta))}{1 + \epsilon^2 +
          2\epsilon\cos(\theta)}\right )
\label{eq:theta}
\ee
\begin{figure}[!htb]
  \begin{center}
    \leavevmode
      \includegraphics[width=7cm, keepaspectratio=true]{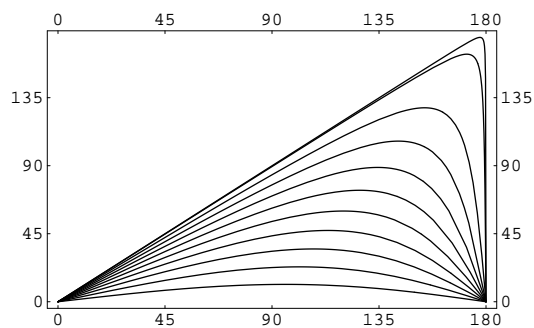}
  \end{center}
  \caption{The amount of shift $\theta-\theta^\prime (\theta ), $ as a function of
  $\theta$ for different values of $\epsilon .$}
\end{figure}
{\bf Remark}: Here $\theta$ is not exactly the "geographical
latitude". It is zero at the "North Pole" ($\sigma_3=+1$), $90$
degrees at the equator, and $180$ degrees at the "South Pole"
($\sigma_3=-1$).

Each map $P(\n ,\epsilon )$ maps the sphere onto itself in an
injective way. For $\n=(0,0,1)$ the map is easy to picture. All
points of the sphere move towards the North Pole along their
meridians, except of the two fixed points: North and South Pole.
All of the Northern hemisphere, and a strip below the equator,
shrinks, while the other part, near the South Pole, stretches.
The amount of stretching can be found by plotting the function
$\theta-\theta^\prime (\theta )$ - it has a maximum at $\theta$
corresponding to $z=-\epsilon$. Thus the parameter $\epsilon$ gets a
simple interpretation: it is the value of $z$ coordinate for which
shrinking of meridians is replaced by stretching - an equilibrium
point. This point is always on the southern hemisphere. For
$\epsilon$ close to zero, where the map is close to the identity
map, the equilibrium point is close to the equator. Then, as
$\epsilon$ approaches the value of $1.0$, corresponding to the
sharp projection operator, the equilibrium latitude gets closer
and closer to the South Pole. In the limit of $\epsilon=1$ all of
the sphere shrinks to the North Pole, only the South Pole remains
where it was.
\subsection{Quantum Fractals and IFS} Our algorithm is, in
fact,  a version of a nonlinear {\sl iterated function system}\,
(IFS).
Such algorithms are known to produce complex geometrical
structures by repeated application of several non-commuting affine
maps. The best known example is the {\sl Sierpinski triangle}\,
generated by random application of $3\times 3$ matrices $A[i],
i=1,2,3$ to the vector:
$v_0=(x_0,y_0,1)$ where $A[i]$ are given by
$A[i]=((0.5,0,ax_i),(0,0.5,ay_i),(0,0,1))$
and $ax_1=1.0, ay_1=1.0, ax_2=1.0, ay_2=0.5, ax_3=0.5, ay_3=1.0.$
(Our $3\times 3$ matrices encode  affine transformations -
usually separated into a $2\times 2$ matrix and a translation
vector.) At each step one of the three transformations $A[i],
i=1,2,3$ is selected with probability $p[i]=1/3$. After each
transformation the transformed vector is plotted on the $(x,y)$
plane. Theoretical papers on IFSs usually assume that the system
is {\sl hyperbolic}\, that is that each transformation is a {\sl
contraction}, i.e. the distances between points get smaller and
smaller. It was shown in \cite{peigne} that this assumption can
be essentially relaxed when transformations are non-linear and act
on a compact space - as is in the case of quantum fractals we are
dealing with. But it is not known whether the results of \cite{peigne}
apply in our case. 
\subsection{Continuous part of the evolution} In our discussion of
 quantum fractals, we neglect completely
the continuous part of the time evolution. Here we are not
interested in "when" jumps happen. We are only interested in the
final pattern produced by a long sequence of jumps. Normally spin
interacts with a magnetic field (if present), which causes our
spin sphere to rotate around the direction of the magnetic field
vector. Such a rotation, between jumps, would smear out our
pattern. To study the pattern we neglect the rotation part. Once
we discard the continuous evolution, the timing of jumps does not
really matter, so we neglect this part of EEQT algorithm (timing
is very important in simulations of particle detectors, arrival
and tunneling times. Here, we simulate jumps as fast as our
computer can crunch the numbers.
\subsection{Nonunitarity}There is one important comment that applies here. Even if we
neglect the continuous time evolution due to magnetic field, the
very presence of the detectors causes a non-unitary time evolution
of the pure state of the quantum system. This evolution is also
called, by some physicists (Dicke, Elitzur, Vaidman) ,
"interaction-free". We do not want to enter into this subject
here, except for one remark: for symmetric geometric
configurations that we are considering here, the EEQT algorithm
implies that this, continuous in time, non-unitary evolution can
be neglected as well. In fact, it follows from the EEQT model that
the "interaction-free" or, as we call it, "binamical part" of the
evolution is determined by the generator $-\kappa\Lambda$, where
$\Lambda=\sum P(\n[i],\epsilon)^2.$ In our case, when our detectors
are symmetrically placed, so that $\sum_{i=1}^N
\n[i]={\bf 0},$ the formula (\ref{eq:pn2}) implies
$\Lambda=\frac{N (1+\epsilon^2)}{4},$ thus the "binamical" part is
just decreasing the norm of the state vector, while leaving its
direction unchanged. Thus it does not affect the geometric pattern
of jumps (it is responsible for the mean frequency of jumps, but
here timing is not important). For a recent review of EEQT,  cf.
\cite{blajaru99}
\subsection{Quantum characteristic exponent}
Averaging our nonlinear PDP over individual histories one gets a
linear Liouville equation for the density matrix of the total
system. Tracing over the classical subsystem is, in our case, easily
performed and we then get:
\be
{\dot \rho}=\kappa (\sum_{i=1}^N P(\n[i],\epsilon)\rho
P(\n[i],\epsilon)-\frac{1}{2}\{\sum_{i=1}^N P(\n[i],\epsilon)^2
,\rho\})
\ee
where $\{\/ ,\/ \}$ stands for anti-commutator, and $\kappa$ is a
coupling constant. For $\rho$ written as
$\rho=\frac{1}{2}(I+\sigma({\bf m}))$, ${\bf
m}^2=m_1^2+m_2^2+m_3^2\leq 1$ after some calculations we get a
very simple time evolution:
$
{\bf m}(t)=\exp \left(-N\kappa\epsilon^2 t/3\right) {\bf
m}(0).$
The quantum characteristic exponent, as defined in Ref
\cite{blajaol99},  is thus $\frac{2N}{3}\kappa$ - not a very
useful quantity in our case. The Hausdorff dimension of the limit
set, for the tetrahedral case, has been numerically estimated in
Ref. \cite{jas} and shown to decrease from 1.44 to 0.49 while
$\epsilon$ increases from 0.75 to 0.95. We hope that by publishing
the generating algorithm we will create interest in confirming
these as well as obtaining new results in this field.
\subsection{How to measure the wave function itself}
\begin{figure}[ht]
\centering
\includegraphics[height=0.40\textheight]
{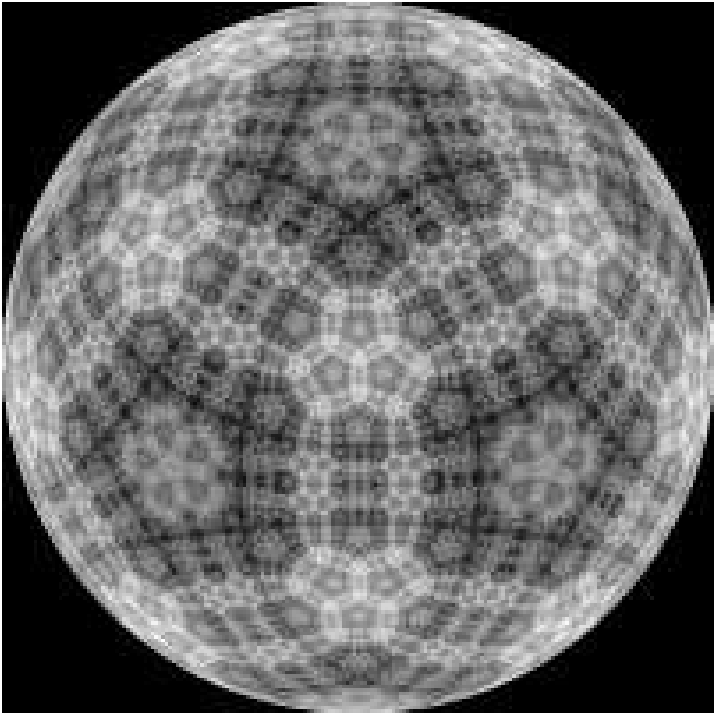}
\caption{Quantum Dodecahedron. $\epsilon=0.78$}\label{fig:dodeca}
\end{figure}

The fractal patterns are produced by a jumping point on the space
of pure quantum states - thus are not directly observable. That is
why we consider our model as simply a toy to play with. But the
model can be developed further on so as to predict observable
effects. One way to do it is by adding another layer of detectors,
densely spaced, that would detect the fractal pattern. Here we
come to the famous question: how to measure the wave function
itself. The idea as to how to model such a measurement within the
EEQT formalism has been indicated in \cite{jad94a}. We hope to
implement the appropriate algorithms at a later time. On the other hand,
interpreting the patterns as resulting from Lorentz boosts applied to 
directions in real space, as discussed in Section \ref{sec:canvas}, we
may try to find similar fractal patterns in the distribution of Galaxies
(compare for instance Figure (\ref{fig:dodeca}) with recent 
paper on the topology of the universe \cite{luminet}.  

\subsection{Detectors} A detector is represented by a two-state
classical system. It can be in one of the two states, denoted $0$
and  $1$. The fact that it is ``classical" means that its two states define
two different superselection sectors that can be mixed statistically, but there
are no observables that connect these sectors.
We will assume that it can "flip" from 0 to 1 or from 1
to 0 when coupled to a quantum spin. Each flip represents an
event; specifically: a detection event. The interpretation is
that when the detector flips, the experimental question "is  the
spin oriented along the vector $\n$?" gets an affirmative answer.
Note that $\n$ and $-\n$ are two different experimental
questions. They corresponds to two opposite spin directions.

A realistic detector should also exhibit a relaxation time, that
is, after each flip it should take some time before it is ready to
flip again. We could easily model this phenomenon in our model,
but here we are interested in patterns that are created, not in
the timing of its appearance .
\subsection{Hyperbolicity\label{sec:hyper}}
In a recent paper \cite{loslzy} {\L}ozinski, S{\l}omczynski and Zyczkowski
studied iterated function systems on the space of mixed states, when
probabilities that are associated with maps are given independently of
the maps. In section III of their paper they give a short discussion
dealing with iterated function systems on the space of pure states as well.
They start with the following definition of a {\sl (pure states) quantum
iterated function system}({\sl QIFS}):

\noindent{\bf Definition (QIFS)} Let ${\go H}_N$ be a complex Hilbert space
of dimension $N.$
Let ${\cal P}_{N}$ be the space of one-dimensional subspaces of ${\go H}_N$.
Given a unit vector $\phi\in {\go H}_N$, let $P_\phi$ be the orthogonal
projection onto the subspace spanned by $\phi.$. Specify two sets of $k$
linear invertible operators:
\begin{itemize}
\item  $V_{i}:{\go H}_{N}\rightarrow {\go H}_{N}$ ($i=1,\ldots ,k$), which
generates maps $F_{i}:{\cal P}_{N}\rightarrow {\cal P}_{N}$ ($i=1,\ldots ,k$
) by
$
\phi\mapsto{V_{i}\phi }/{\| V_{i}
\phi \| }
$
for any $\phi \in {\cal P}_N$, and

\item  $W_{i}:{\go H}_{N}\rightarrow {\go H}_{N}$ ($i=1,\ldots ,k$),
forming an operational resolution of identity,
$\sum_{i=1}^{k}W_{i}^{\dagger }W_{i}={\mathbb I}$, which generates
probabilities $p_{i}:{\cal P}_{N}\rightarrow \left[ 0,1\right] $ ($
i=1,\ldots ,k$) by
$
p_{i}\left(\phi\right) :=\left\| W_{i}\left( \phi \right) \right\| ^{2} \ .
\label{probabilities}
$
\end{itemize}

\noindent{\bf Comments}
\begin{enumerate}
\item The authors define the system to be hyperbolic if the maps $F_i$ are
contractions with respect to the Fubini-Study distance
$d(P_\phi,P_\psi )= \arccos \left(\sqrt{Tr (P_\phi P_\psi
)}\right),$ i.e. there exists constants $0< L_i < 1$ such that
$d(F_i(P_\phi ),F_i(P_\psi)) \le L_i d(P\phi ,P_\psi )$ for all
$\phi , \psi \in {\go H}_N.$ Then they state a proposition
(Proposition 1 in \cite{loslzy}) that guarantees existence of an
invariant measure for a hyperbolic system. It seems that the
assumptions of this proposition can not be satisfied. It is well
known   \cite{greub1} that a smooth injective map of a compact
orientable manifold is automatically a bijection. On the other
hand, iterating the map several times if necessary, the distance
between any two image points can make smaller than any given
positive number, and therefore, {\sl a fortiori}\, less than the
maximal distance between the points of  ${\cal P}_N$, which
contradicts surjectivity.\footnote{After informing the authors
about this problem, they kindly replied that they saw it too, that
their {\bf Proposition 1} has been deleted from the paper they
submitted to Physical Review, and that they are going to replace
the electronic version of the paper on the eprint server as well.}
\item The authors of \cite{loslzy} assume that the probabilities $p_i$ are
given independently of the mapping operators $V_i.$ This is not the case
in the EEQT scheme. In EEQT $V_i$ need not be invertible, but the probabilities
$p_i$ are determined by the $V_i$-s automatically, and in such a way that
whenever there is a danger of dividing by zero, the associated probability
is automatically zero.
\end{enumerate}
\ack{ One of us (A.J) would like to thank L.K-J
for invaluable help, and to Palle Jorgensen for encouragement and useful comments.}

\begin{figure}[ht]
\centering
\includegraphics[height=0.40\textheight]
{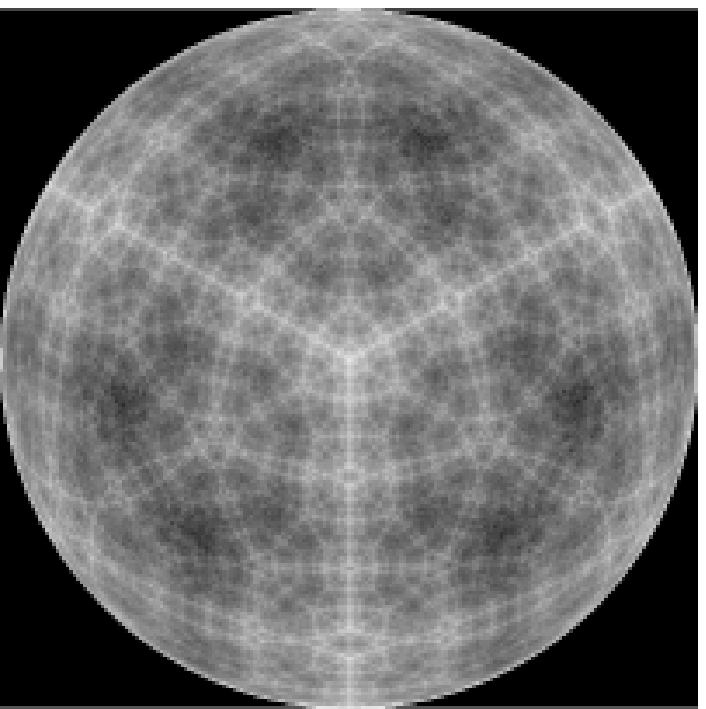}
\caption{Quantum Tetrahedron. $\epsilon=0.5$ This is the simplest case. \\The idea has been first described in
\cite{jad93e} and then exploited in \cite{jas}.}
\end{figure}
\pagebreak
\begin{figure}[ht]
\centering
\includegraphics[height=0.40\textheight]
{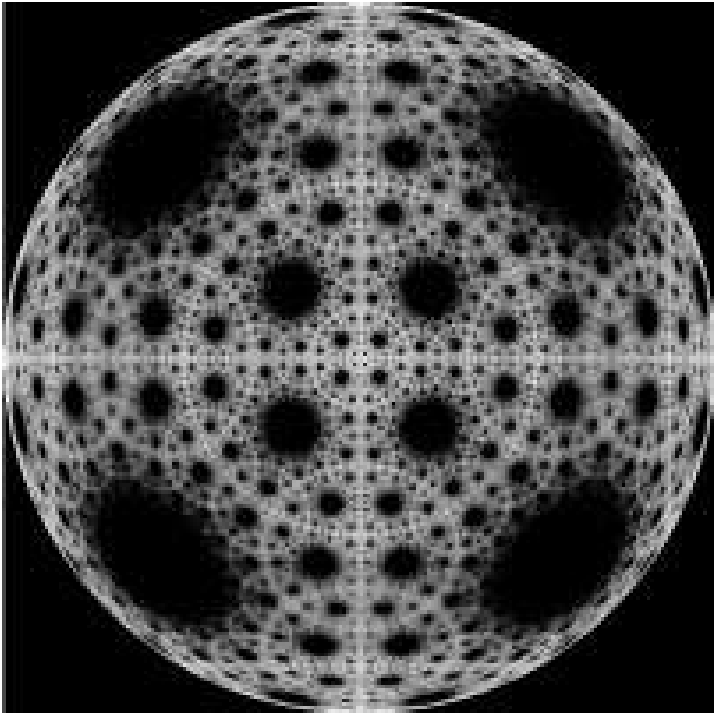}
\caption{Quantum Octahedron. $\epsilon=0.58$}
\end{figure}
\begin{figure}[ht]
\centering
\includegraphics[height=0.40\textheight]
{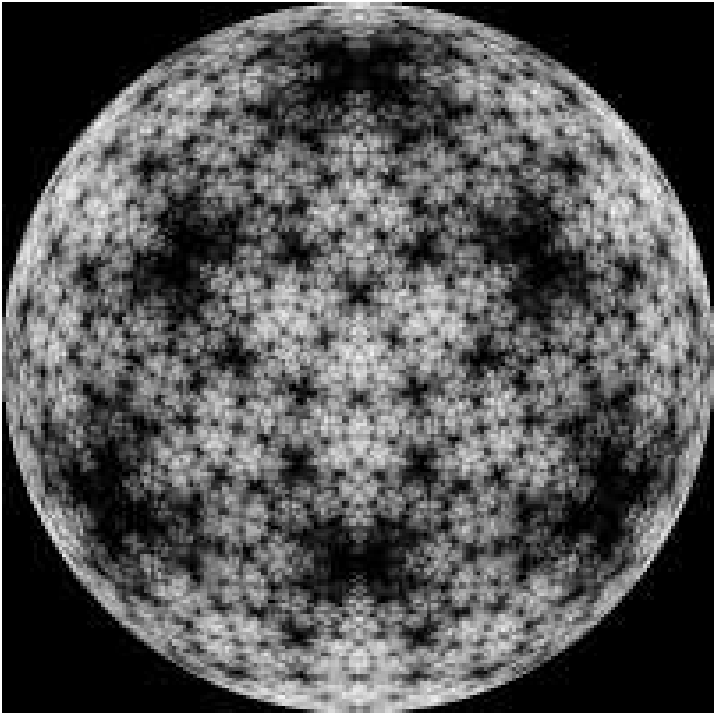}
\caption{Quantum Cube. $\epsilon=0.7$}
\end{figure}
\pagebreak
\begin{figure}[ht]
\centering
\includegraphics[height=0.40\textheight]
{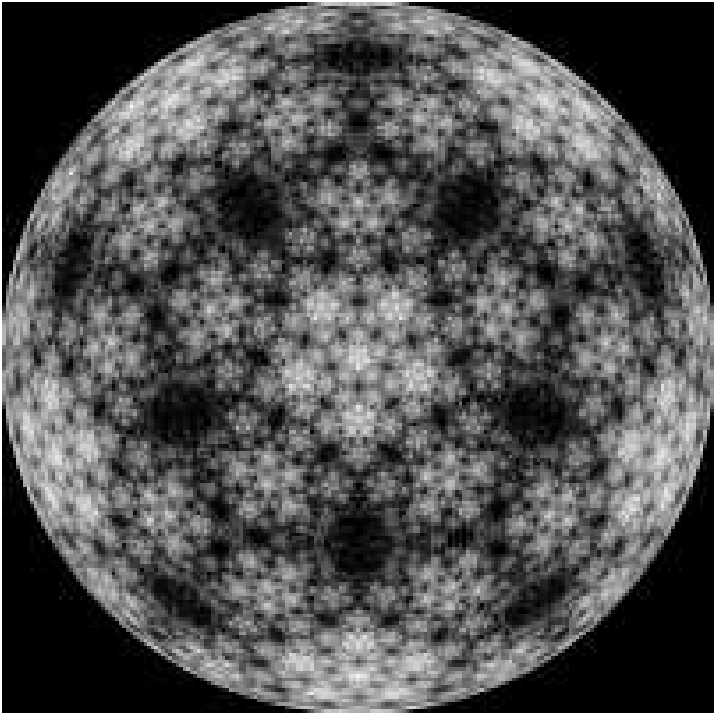}
 \caption{Quantum Icosahedron. $\epsilon=0.75$}
\end{figure}
\pagebreak
\begin{figure}[ht]
\centering
\includegraphics[height=0.40\textheight]
{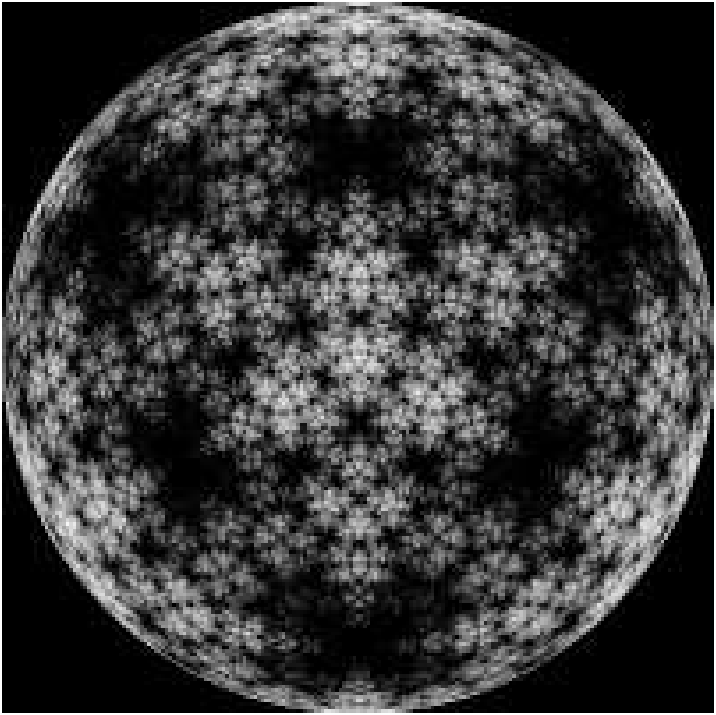}
\caption{Quantum Double Tetrahedron. $\epsilon=0.7$}
\end{figure}
\begin{figure}[ht]
\centering
\includegraphics[height=0.40\textheight]
{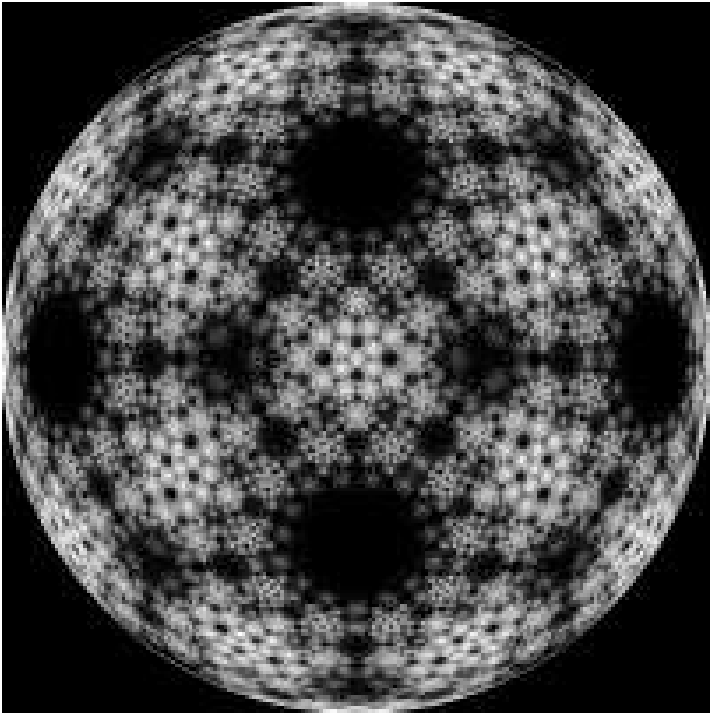}
\caption{Quantum Icosidodecahedron. $\epsilon=0.85$}
\end{figure}\end{document}